\documentclass[aps, prl, tightenlines, floatfix, twocolumn, amsmath, superscriptaddress, showkeys]{revtex4-1}
\usepackage{float}
\usepackage{graphicx}
\usepackage{dcolumn}
\usepackage{bm}
\usepackage{bbold}
\usepackage{enumerate}
\usepackage[utf8]{inputenc}
\usepackage[english]{babel}
\usepackage{amsthm}

\usepackage{xcolor}

\newtheorem{theorem}{Theorem}[section]

\newtheorem{lemma}[theorem]{Lemma}

\begin{document}

\title{Precision cosmology made more precise}

\author{Giorgio Galanti}
\email{gam.galanti@gmail.com}
\affiliation{INAF, Istituto di Astrofisica Spaziale e Fisica Cosmica di Milano, Via A. Corti 12, I -- 20133 Milano, Italy}

\author{Marco Roncadelli}
\email{marco.roncadelli@pv.infn.it}
\affiliation{INFN, Sezione di Pavia, Via A. Bassi 6, I -- 27100 Pavia, Italy}
\affiliation{INAF, Osservatorio Astronomico di Brera, Via E. Bianchi 46, I -- 23807 Merate, Italy}

\date{\today}

\begin{abstract}
So far, the standard attitude to solve the Friedmann equations in the simultaneous presence of radiation $R$, matter $M$ and cosmological constant ${\Lambda}$ is to find solutions $R_R (t)$, $R_M (t)$ and $R_{\Lambda} (t)$ {\it separately} for each {\it individual  component alone}, and next to join them together, thereby  obtaining a {\it piecewise solution} $R_{\rm pw} (t)$. We instead find the {\it exact} and {\it analytic} solution $R (t)$ of the same equations in flat space. Moreover, we quantify the error made when $R_{\rm pw} (t)$ is used in place of $R (t)$.
\end{abstract}



\maketitle


{\it Introduction} -- A wide spread belief is that cosmology has entered its precision phase. Of course, we still do not know the nature of non-baryonic dark matter and dark energy -- represented here by the cosmological constant $\Lambda$ -- and the cosmological parameters $H_0$, $\Omega_{R,0}$, $\Omega_{M,0}$ and $\Omega_{\Lambda}$ are not determined very precisely~\cite{note1}. But when everything is put together a very remarkably consistent scenario emerges.

Nevertheless, the Friedmann equations are still solved in a too naive fashion. Basically, their solutions $R_R (t)$, $R_M (t)$ and $R_{\Lambda} (t)$ are found {\it separately} for each {\it individual component} $\Omega_R (t)$, $\Omega_M (t)$ and $\Omega_{\Lambda}$, respectively, {\it alone}. And next $R_R (t)$, $R_M (t)$ and $R_{\Lambda} (t)$ are joined together, thereby providing a {\it piecewise solution} $R_{\rm pw} (t)$. Such a procedure looks correct whenever one component dominates over the others, so that effectively only a single component is relevant. Yet, we know that there is a time $t_{M \Lambda}$ when $\Omega_M (t_{M \Lambda}) = \Omega_{\Lambda}$ and a previous time $t_{R M}$ when $\Omega_R (t_{R M}) = \Omega_M (t_{R M})$. Around $t_{M \Lambda}$ and $t_{R M}$ two components contribute almost equally to the energy budget of the Universe: in such a situation the piecewise approximation manifestly breaks down. In addition, being the Friedmann equations non-linear in $R (t)$, the contribution of, say, 
$\Omega_M (t)$ plus $\Omega_{\Lambda}$ gives rise to a scale factor $R_{M \Lambda} (t)$ {\it different} from the sum of the scale factors $R_M (t)$ and $R_{\Lambda} (t)$ as computed by taking $\Omega_M (t)$ {\it alone} and $\Omega_{\Lambda}$ {\it alone}, respectively, into account. In some textbook the behavior of the Universe in the presence of two components is discussed rather cursorily  (see e.g.~\cite{ryden}), but -- to the best of our knowledge -- so far the scale factor $R (t)$ has never been computed exactly and analytically in the simultaneous presence of $\Omega_R (t)$, $\Omega_M (t)$ and 
$\Omega_{\Lambda}$ within the Friedmann equations. 

The aim of the present Letter is to fill this gap, and to discuss some of its implications.

{\it Notations and conventions} -- It is more useful to work with the normalized scale factor $a (t) \equiv R (t)/R_0$ (with $R_0 \equiv R (t_0)$), so that the Friedmann 
equations are
\begin{equation}
\label{Friedmann}
\left( \frac{\dot{a}(t)}{a(t)} \right)^2=\frac{8\pi G}{3 c^2}\epsilon(t)-\frac{k c^2}{R_0^2a^2(t)}~,
\end{equation}
\begin{equation}
\label{FriedmannII}
\frac{\ddot{a}(t)}{a(t)} = -\frac{4 \pi G}{3 c^2}\Big( \epsilon(t)+3 P(t) \Big)~,
\end{equation}
where $G$ is the Newton gravitational constant, $c$ is the speed of light {\it in vacuo}, $k$ 
is the curvature constant, while $\epsilon(t)$ represents the total energy density of the cosmic fluid \begin{equation}
\label{epsilon}
\epsilon(t)=\frac{3 c^2 H_0^2}{8 \pi G}\left(\frac{\Omega_{R,0}}{a^4(t)}+\frac{\Omega_{M,0}}{a^3(t)}+\Omega_{\Lambda} \right)~,
\end{equation}
and $P(t)$ denotes the pressure of the cosmic fluid 
\begin{equation}
\label{press}
P(t)=\frac{c^2 H_0^2}{8 \pi G} \left(\frac{\Omega_{R,0}}{a^4(t)} - 3\Omega_{\Lambda} \right)~.
\end{equation}

We henceforth restrict our attention to a metrically flat there-dimensional space ($k = 0$), for which there is nowadays a general consensus. Accordingly, by inserting Eq.~(\ref{epsilon}) into Eq.~(\ref{Friedmann}) with $k = 0$, and Eqs.~(\ref{epsilon}) and~(\ref{press}) into Eq.~(\ref{FriedmannII}), we end up with
\begin{equation}
\label{Friedmann1}
\left( \frac{\dot{a}(t)}{a(t)} \right)^2=H_0^2\left(\frac{\Omega_{R,0}}{a^4(t)}+\frac{\Omega_{M,0}}{a^3(t)}+\Omega_{\Lambda} \right)~,
\end{equation}
\begin{equation}
\label{Friedmann1II}
\frac{\ddot{a}(t)}{a(t)} =-H_0^2 \left( \frac{\Omega_{R,0}}{a^4(t)}+\frac{1}{2}\frac{\Omega_{M,0}}{a^3(t)} - \Omega_{\Lambda} \right)~.
\end{equation}
Moreover, by exploiting Eq.~(\ref{epsilon}), we find that the scale factor when radiation and matter contribute equally reads
\begin{equation}
\label{aRM}
a_{R M} \equiv a(t_{R M}) = \frac{\Omega_{R,0}}{\Omega_{M,0}}~,
\end{equation}
while the scale factor when the matter and dark energy contributions are equal is given by
\begin{equation}
\label{aML}
a_{M \Lambda} \equiv a (t_{M \Lambda}) = \left( \frac{\Omega_{M,0}}{\Omega_{\Lambda}} \right)^{1/3}~.
\end{equation}

{\it Piecewise solution} -- As a preliminary step, we briefly recall the piecewise solution  $a_{\rm pw} (t)\equiv R_{\rm pw} (t)/R_0$ usually found in the literature -- describing separately the three epochs of radiation, matter and cosmological constant do\-mi\-nation -- since it provides a benchmark for comparison with our exact solution. By solving Eq.~(\ref{Friedmann1}) in the single-component case, we find 
\begin{equation}
\label{FriedmannPiecewise}
a_{\rm pw}(t)=
\begin{cases}
K_R t^{1/2} &   \,\, t \le \tilde{t}_{R M}~,\\[8pt]

K_M t^{2/3} &   \,\, \tilde{t}_{R M} < t \le \tilde{t}_{M \Lambda}~,\\[8pt]

K_{\Lambda} {\rm exp } \left( \Omega_{\Lambda}^{1/2} H_0 t \right) &   \,\, t > \tilde{t}_{M \Lambda}~,
\end{cases}
\end{equation}
where $K_R$, $K_M$ and $K_{\Lambda}$ are normalization constants, and $\tilde{t}_{R M}$ and $\tilde{t}_{M \Lambda}$ have the same conceptual meaning of $t_{R M}$ and $t_{M \Lambda}$, respectively~\cite{note2}. We obtain $K_{\Lambda}$ by imposing the condition $a=1$ today, so that we get
\begin{equation}
\label{KL}
K_{\Lambda}={\rm exp } \left( - \Omega_{\Lambda}^{1/2} H_0 t_0 \right)~,
\end{equation}
while $K_M$ and $K_R$ are fixed by requiring continuity of $a (t)$ at the junction points 
$\tilde{t}_{M\Lambda}$ and $\tilde{t}_{RM}$. Because the latter two quantities are still unknown, we turn the argument around by demanding continuity of the inverse function $t_{\rm pw}(a)$ (which is always single-valued since $a (t)$ is monotonically increasing)
\begin{equation}
\label{FriedmannPiecewiseInv}
t_{\rm pw}(a)=
\begin{cases}
K_R^{-2} a^2 &   \,\, a \le a_{RM}~,\\[8pt]

K_M^{-3/2} a^{3/2} &   \,\, a_{RM} < a \le a_{M\Lambda}~,\\[8pt]

t_0+\Omega_{\Lambda}^{-1/2}H_0^{-1} \, {\rm ln} \, a &   \,\, a > a_{M\Lambda}~.
\end{cases}
\end{equation}
Specifically, from the second and third of Eq. (\ref{FriedmannPiecewiseInv}) we find    $K_M^{-3/2} a_{M\Lambda}^{3/2} = t_0+\Omega_{\Lambda}^{-1/2}H_0^{-1} \, {\rm ln} \, a_{M\Lambda}$, which implies  
\begin{equation}
\label{KM}
K_M = \left( \frac{\Omega_{M,0}}{\Omega_{\Lambda}} \right)^{1/3} \left[ H_0 t_0 + \frac{1}{3} \Omega_{\Lambda}^{-1/2} {\rm ln} \left(\frac{\Omega_{M,0}}{\Omega_{\Lambda}}\right) \right]^{-2/3} H_0^{2/3}~,
\end{equation}
where we have employed Eq.~(\ref{aML}) with $t_{M \Lambda} \to \tilde{t}_{M \Lambda}$. Similarly, from the first and second of Eq. (\ref{FriedmannPiecewiseInv}) we further obtain $K_R^{-2} a_{RM}^2 = K_M^{-3/2} a_{RM}^{3/2}$, which entails
\begin{equation}
\label{KR}
K_R= \left(\frac{\Omega_{R,0}}{\Omega_{\Lambda}} \right)^{1/4} \left[ H_0 t_0 + \frac{1}{3} \Omega_{\Lambda}^{-1/2} {\rm ln} \left(\frac{\Omega_{M,0}}{\Omega_{\Lambda}} \right) \right]^{-1/2} H_0^{1/2}~,
\end{equation}
where now we have used Eq.~(\ref{aRM}) with $t_{R M} \to \tilde{t}_{R M}$.

What remains to be done is the determination of $\tilde{t}_{R M}$ and $\tilde{t}_{M \Lambda}$. Being by definition $\tilde{t}_{R M} \equiv t_{\rm pw}(a_{R M})$, thanks to the first of Eq. (\ref{FriedmannPiecewiseInv}) combined with Eqs. (\ref{aRM}) and (\ref{KR}) we get
\begin{equation}
\label{ttRM}    
\tilde{t}_{RM}=\Omega_{R,0}^{3/2} \Omega_{M,0}^{-2} \Omega_{\Lambda}^{1/2} \left[t_0 + \frac{1}{3} \Omega_{\Lambda}^{-1/2} H_0^{-1} {\rm ln} \left(\frac{\Omega_{M,0}}{\Omega_{\Lambda}}\right) \right]~.
\end{equation}
Moreover -- since by definition $\tilde{t}_{M \Lambda} \equiv t_{\rm pw}(a_{M \Lambda})$ -- owing to the second of Eq. (\ref{FriedmannPiecewiseInv}) in conjunction with Eqs. (\ref{aML}) and (\ref{KM}) we find
\begin{equation}
\label{ttML}
\tilde{t}_{M \Lambda}=t_0 + \frac{1}{3} \Omega_{\Lambda}^{-1/2} H_0^{-1} {\rm ln} \left(\frac{\Omega_{M,0}}{\Omega_{\Lambda}}\right)~.
\end{equation}
With the benchmark values of the numerical parameters reported in~\cite{note3} we have 
$\tilde{t}_{R M} = 6.34 \times 10^4 \, {\rm yr}$ and $\tilde{t}_{M \Lambda} = 8.75 \, {\rm Gyr}$.

{\it Our solution} -- We start by rewriting Eq.~(\ref{Friedmann1}) as 
\begin{equation}
\label{Friedmann2}
H_0^{-1} \frac{{\rm d}a(t)}{{\rm d}t}=\left( \frac{\Omega_{R,0}}{a^2(t)}+\frac{\Omega_{M,0}}{a(t)}+\Omega_{\Lambda}a^2(t) \right)^{1/2}~,
\end{equation} 
and by performing the separation of variables, it takes the integral form
\begin{equation}
\label{Friedmann3}
H_0 t=\int_{0}^{a(t)} {\rm d}a' \left( \frac{\Omega_{R,0}}{a'^2} +\frac{\Omega_{M,0}}{a'}+ \Omega_{\Lambda}a'^2 \right)^{-1/2}~,
\end{equation}
which allows us to express $t$ as a function of $a$. Our ultimate goal is to have instead $a$ as a function of $t$.

A direct attempt at inverting Eq.~(\ref{Friedmann3}) is extremely cumbersome. We can save work by splitting up such an equation -- which contains three $\Omega$ parameters -- into a pair of equations, each including only two of them. This task can be accomplished  as follows. Neglecting primordial inflation, standard cosmology tells us that the Universe is radiation dominated for $0 \leq t \leq t_{\rm R M}$, matter dominated for $t_{\rm R M} < t \leq t_{{\rm M} \Lambda}$ and dark energy dominated for $t_{{\rm M} \Lambda} < t \leq t_0$. So, we can define a time $t_s$ such that $t_{R M} \ll t_s \ll t_{{\rm M} \Lambda}$ (more about this choice, later). Being $t_s$ well inside the regime of matter domination, it is sure that for $t \leq t_s$ only radiation or matter dominates, whereas for $t > t_s$ only matter or dark energy dominates. As a consequence, Eq.~(\ref{Friedmann3}) becomes equivalent to the two equations

\begin{widetext}

\begin{equation}
\label{Friedmann4}
H_0 t=
\begin{cases}
\displaystyle\int_{0}^{a(t)} {\rm d}a' \left(  \dfrac{\Omega_{R,0}}{a'^2} +\dfrac{\Omega_{M,0}}{a'} \right)^{-1/2}~, & a \leq a_s~,\\[8pt]
H_0 t_s + \displaystyle\int_{a_s}^{a(t)} {\rm d}a' \left(  \dfrac{\Omega_{M,0}}{a'}+ \Omega_{\Lambda}a'^2 \right)^{-1/2}~, & a > a_s~,
\end{cases}
\end{equation}
where $a_s \equiv a(t_s)$. The solution of Eq.~(\ref{Friedmann4}) is

\begin{equation}
\label{Friedmann5}
H_0 t=
\begin{cases}
\displaystyle \dfrac{2}{3\Omega_{M,0}^2}\left[ (\Omega_{M,0} a - 2 \Omega_{R,0})(\Omega_{M,0} a+\Omega_{R,0})^{1/2} + 2 \Omega_{R,0}^{3/2}  \right]~, & a \leq a_s~,\\[8pt]
H_0 t_s + \displaystyle \dfrac{2}{3\Omega_{\Lambda}^{1/2}}{\rm ln}\left[\frac{ \Omega_{\Lambda} a^{3/2} + \Omega_{\Lambda}^{1/2} \left( \Omega_{M,0} +\Omega_{\Lambda}a^3 \right)^{1/2}}{ \Omega_{\Lambda} a_s^{3/2} + \Omega_{\Lambda}^{1/2} \left( \Omega_{M,0}+\Omega_{\Lambda}a_s^3 \right)^{1/2}}\right]~, & a > a_s~.
\end{cases}
\end{equation}

\end{widetext}

In order to proceed towards our goal -- and so getting the explicit function $a (t)$ -- we turn both expressions in Eq.~(\ref{Friedmann5}) into two third-order algebraic equations. Then, the fundamental theorem of algebra ensures that they have three roots, which can be either all real or one real and two complex conjugate, since the equations in question have real coefficients. The calculation of $a (t)$ in the second of Eq.~(\ref{Friedmann5}) is rather straightforward, and by inspection one can see that we are in the case of one real and two complex conjugate roots. Being the latter ones physically unacceptable, the behavior of $a (t)$ for $t > t_s$ is reported in the third line of Eq.~(\ref{FriedmannSolution}) below.

The calculation of $a (t)$ in the first of Eq.~(\ref{Friedmann5}) is instead not straightforward at all, and deserves a detailed discussion which is reported in the Supplementary Material (SM). Accordingly, we find the behavior of $a (t)$ for $t \le t_s$ as reported in the first and second line of Eq.~(\ref{FriedmannSolution}) below.

Thus, our {\it full exact solution} $a (t)$ of the first Friedmann equation~(\ref{Friedmann1}) is

\begin{widetext}
\begin{equation}
\label{FriedmannSolution}
a(t)=
\begin{cases}

\begin{aligned}[]
\frac{\Omega_{R,0}}{\Omega_{M,0}}\left\{1-2 \, {\rm sin}\left[\frac{1}{3} \, {\rm arcsin}\left( 1-3\frac{\Omega_{M,0}^2}{\Omega_{R,0}^{3/2}}H_0 t+\frac{9}{8}\frac{\Omega_{M,0}^4}{\Omega_{R,0}^3}H_0^2t^2 \right) \right]\right\}
\end{aligned}~, &   \,\, 0 \leq t \leq t_*~,\\[8pt]

\\

\\

\begin{aligned}[]
 \frac{\Omega_{R,0}}{\Omega_{M,0}}\left\{1+2 \, {\rm cos}\left[\frac{1}{3} \, {\rm arccos}\left( 1-3\frac{\Omega_{M,0}^2}{\Omega_{R,0}^{3/2}}H_0 t+\frac{9}{8}\frac{\Omega_{M,0}^4}{\Omega_{R,0}^3}H_0^2t^2 \right) \right]\right\}
\end{aligned}~, &   \,\, t_* < t \le t_s~,\\[8pt]

\\

\\

\begin{aligned}[]

\left\{ a_s^{3/2} {\rm cosh}\left[ \frac{3}{2} \Omega_{\Lambda}^{1/2} H_0 (t-t_s) \right]+ \left( a_s^3+\frac{\Omega_{M,0}}{ \Omega_{\Lambda}} \right)^{1/2} {\rm sinh}\left[ \frac{3}{2} \Omega_{\Lambda}^{1/2} H_0 (t-t_s)\right] \right\}^{2/3}

\end{aligned}~, &  \,\, t > t_s~,

\end{cases}
\end{equation}

\end{widetext}
where we have set $t_* \equiv 4 \Omega_{R,0}^{3/2}/(3 \Omega_{M,0}^2 H_0)$ (see also the SM), and $a_s$ is calculated by employing the second of Eq.~(\ref{FriedmannSolution}) with $t=t_s$.

It can be checked that Eq.~(\ref{FriedmannSolution}) is solution of Eq.~(\ref{Friedmann1}) by taking $\Omega_{\Lambda}=0$ for $t \le t_s$ (where only radiation and matter are important) and $\Omega_R (t)=0$ for $t > t_s$ (where only matter and dark energy are relevant). And by employing these prescriptions, it is possible to see that Eq.~(\ref{FriedmannSolution}) is solution also of the second Friedmann equation~(\ref{Friedmann1II}) even though this is unnecessary, since such an equation can be replaced by the conservation equation~\cite{notaM}
\begin{equation}
\label{09012021a}
\dot{\epsilon} (t) + 3 \bigl(\epsilon (t) + P (t) \bigr) \frac{\dot{a}(t)}{a(t)} = 0~,
\end{equation}
which is {\it identically} satisfied thanks to Eqs. (\ref{epsilon}) and (\ref{press}). Thus, Eq.~(\ref{FriedmannSolution}) describes the {\it whole evolution} of $a (t)$ in the present situation.

{\it Discussion} -- We are now ready to investigate the properties of our exact solution $a (t)$. In order to find out the behavior of $a$ as a function of $H_0 t$, we take the benchmark values of the relevant parameters reported in~\cite{note3}. Accordingly, in the upper panel of Figure~\ref{SF} we plot $a$ versus $H_0 t$, along with the asymptotic behavior as given by $a_{\rm pw}$.  
\begin{figure}[h]
\centering
\includegraphics[width=0.48\textwidth]{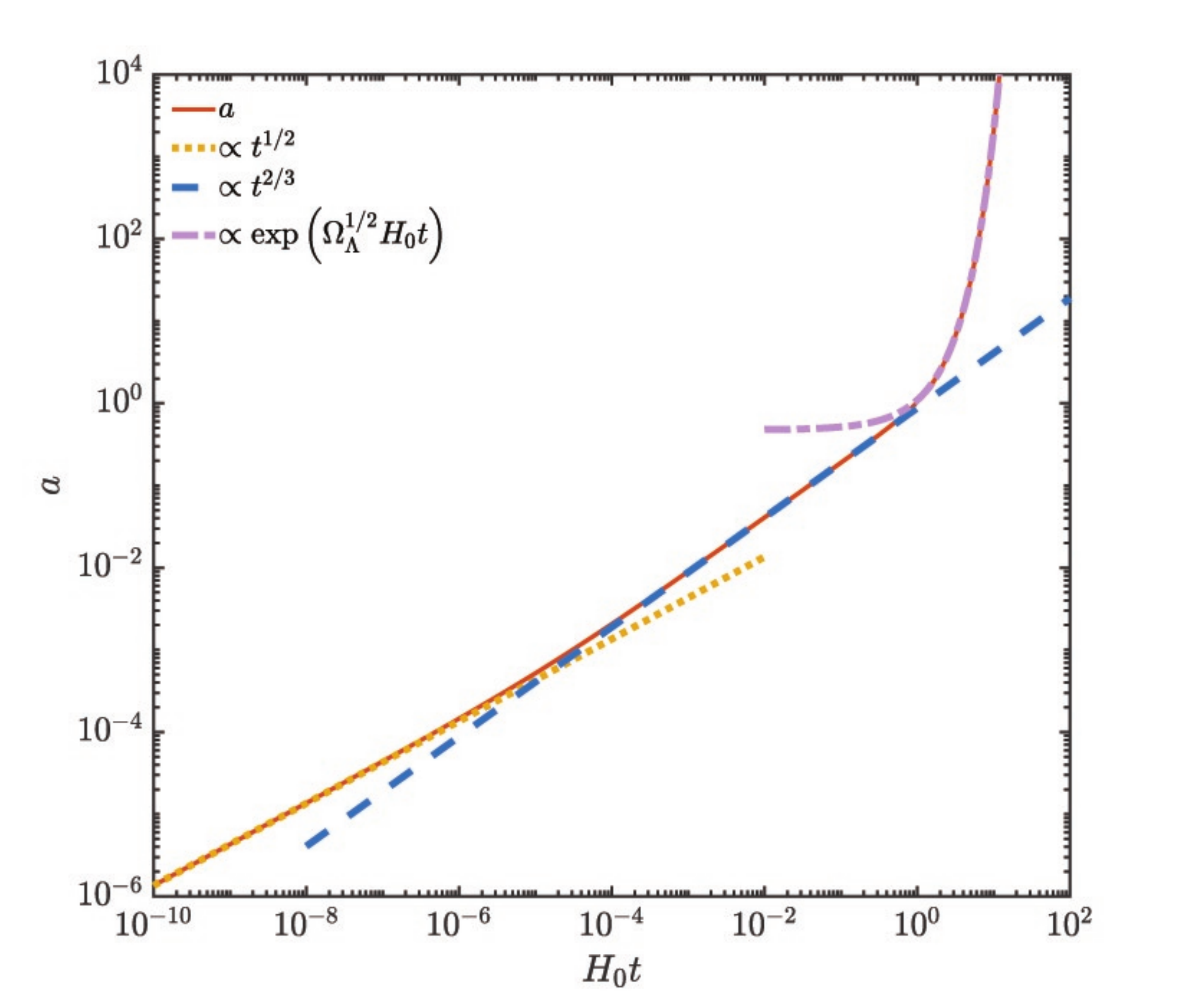}
\includegraphics[width=0.48\textwidth]{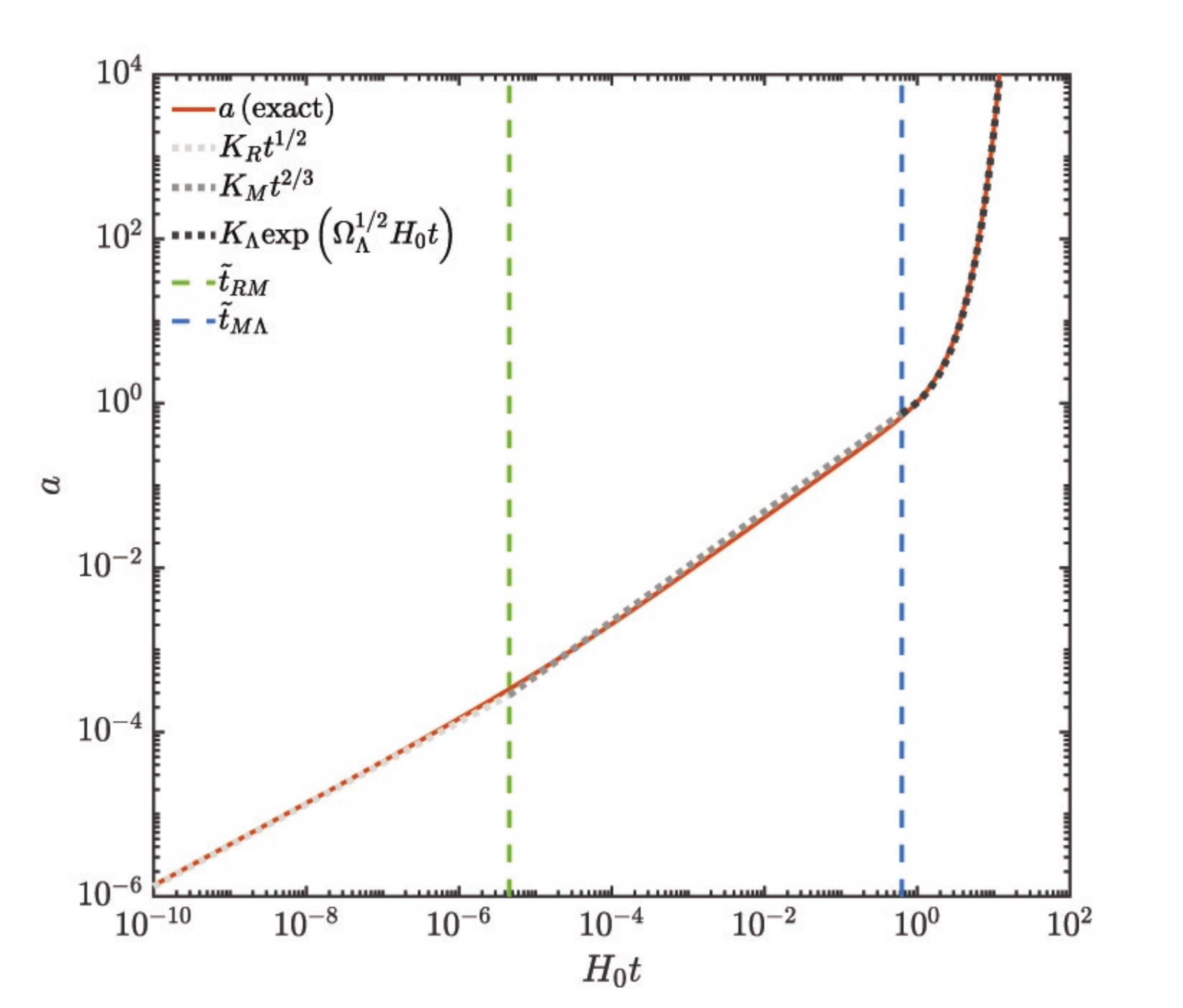} 
\caption{\label{SF} {\it Upper panel}: using our benchmark values, we plot the scale factor $a$ and the asymptotic behaviors given by $a_{\rm pw}$ versus $H_0t$. {\it Lower panel}:  exact solution $a$ and  piecewise one $a_{\rm pw}$ as plotted versus $H_0t$. For the reader's convenience, we report this Figure in a larger format in the SM.} 
\end{figure} 
Moreover, thanks to Eq. (\ref{aRM}) and the first of Eq.~(\ref{Friedmann5}),  $t_{RM}$ is provided by
\begin{equation}
\label{tRM}
t_{R M}=\frac{4 \Omega_R^{3/2}}{3\Omega_M^2 H_0}\left( 1-2^{-1/2}\right)~,
\end{equation}
while on account of Eq. (\ref{aML}) and of the second of Eq.~(\ref{Friedmann5}) $t_{M\Lambda}$ can be written as
\begin{equation}
\label{t M L}
t_{M \Lambda}=t_s+\frac{2}{3\Omega_{\Lambda}^{1/2}H_0}{\rm ln}\left[\frac{ \Omega_{M,0}^{1/2}  \left( 1+ 2^{1/2} \right) }{ \Omega_{\Lambda}^{1/2} a_s^{3/2} +  \left( \Omega_{M,0}+\Omega_{\Lambda}a_s^3 \right)^{1/2}}\right]~.
\end{equation}
Superficially, Eq. (\ref{t M L}) might look suspicious owing to the dependence of $t_{M \Lambda}$ on $t_s$ and $a_s$ -- given the freedom to arbitrarily select $t_s$ within the rather wide range $t_{R M} \ll t_s \ll t_{{\rm M} \Lambda}$ -- which correspondingly fixes $a_s$. The best way to clarify this point amounts to recall that Eq. (\ref{t M L}) is the solution of the Friedmann equations -- in the form of the second of Eq. (\ref{Friedmann5}) -- specialized to the case $a = a_{M \Lambda}$ according to Eq. (\ref{aML}). As we shall shortly see, a very good choice is $t_s = (t_{R M} \, t_0)^{1/2}$. Correspondingly, from Eqs. (\ref{tRM}) and (\ref{t M L}) we find $t_{R M} = 4.73 \times 10^4 \, \rm yr$, $t_s = 25.23 \, {\rm Myr}$, $a_s = 0.013$ and $t_{M \Lambda} = 9.82 \, \rm Gyr$. So, $t_s$ indeed turns out to be just halfway between $t_{R M}$ and $t_{M \Lambda}$, and in addition $t_* = 1.61 \times 10^5 \, {\rm yr}$, thereby implying $t_* \ll t_s$ as previously assumed. 

{\it Conclusions} -- We have derived the exact analytic solution of the Friedmann equations in the presence of radiation, matter and cosmological constant as time increases, assuming that three-dimensional space is metrically flat. We close this Letter by comparing our solution with the piecewise one. This is best done by plotting both of them in the lower panel of Figure~\ref{SF} versus $H_0t$. From that we see that $a_{\rm pw} (t)$ roughly approximates $a$ at any time. Specifically, as $t$ increases in the radiation dominated epoch $a_{\rm pw} (t)$  underestimates $a (t)$, while in the matter dominated era $a_{\rm pw} (t)$ crosses $a (t)$, thus first underestimating and then overestimating the exact solution. Only inside the dark energy dominated epoch appears $a_{\rm pw} (t)$ as a good approximation of $a$. In order to quantify the accuracy of the piecewise solution we estimate the error with respect to the exact solution~(\ref{FriedmannSolution}) in Table~\ref{error}, where $t_M$ is a generic time when the Universe is matter dominated. Finally, note that $t_{R M}/\tilde{t}_{R M} = 0.75$ and $t_{M \Lambda}/\tilde{t}_{M \Lambda} = 1.12$. 

\

\begin{table}[H]
\begin{center}
\begin{tabular}{c|c}          
\hline
$t$  & $(a_{\rm pw} (t) - a (t))/a (t) \, [\%]$  \\
\hline
\hline
$0.01 \, t_{RM}$ &    $-3.9$  \\
$t_{RM}$ &    $-13.5$  \\
$\tilde{t}_{RM}$ &    $-14.9$  \\
$t_M$ &    $-14.9 \div 15.3$  \\
$\tilde{t}_{M\Lambda}$ &    $9.8$  \\
$t_{M\Lambda}$ &    $6.6$  \\
$t_0$ &    $0$  \\
\hline
\end{tabular}
\caption{Relative error of $a_{\rm pw} (t)$ with respect to $a (t)$ at different cosmic times. The error at $t_0$ is zero by construction.}  
\label{error}
\end{center}    
\end{table} 
Thus, the largest discrepancy between $a_{\rm pw} (t)$ and $a (t)$ occurs in the matter dominated regime even around $t_{RM}$, and next progressively decreases at smaller and smaller times.

\section*{Acknowledgments}

G. G. acknowledges contribution from the grant ASI-INAF 2015-023-R.1, while the work of M. R. is supported by an INFN TAsP grant.

\newpage

\begin{widetext}

\appendix

\centerline{{\bf SUPPLEMENTARY MATERIAL}}

\section{Solution of the first of Eq.~(19)}

For the reader's convenience, we report below the first of Eq.~(19) 
\begin{eqnarray}
\label{eqRM}
H_0 t =  \frac{2}{3\Omega_{M,0}^2}\left[ (\Omega_{M,0} \, a - 2 \Omega_{R,0})(\Omega_{M,0} \, a + \Omega_{R,0})^{1/2} + 2 \Omega_{R,0}^{3/2}  \right]~, \ \ \ \ \ \ \ \ \ a \le a_s~,
\end{eqnarray}
where $a_s \equiv a(t_s)$ is defined in the main text. Our aim is to study its solutions and their behavior in order to single out the physical one. To this end, we define the polynomial
\begin{eqnarray}
\label{poly} 
P(a,t) \equiv 4 \Omega_{M,0}^3 \, a^3 -12 \Omega_{M,0}^2 \Omega_{R,0} \, a^2 
 - 9 \Omega_{M,0}^4 H_0^2 t^2 + 24  \Omega_{M,0}^2 \Omega_{R,0}^{3/2} H_0 t~,
\end{eqnarray}
where $a$ should be regarded as a variable while $t$ as a parameter. In view of our subsequent needs, we also quote its derivative
\begin{eqnarray}
\label{polyD} 
\frac{{\rm d}P(a,t)}{{\rm d} a} = 12 \Omega_{M,0}^3 \, a^2 - 24 \Omega_{M,0}^2 \Omega_{R,0} \, a~.
\end{eqnarray}
Hence, we have the following:

\begin{lemma}
\label{lemma1}
$P(a, t)$ is a decreasing function of $a$ for $0<a<2\Omega_{R,0}/\Omega_{M,0}$ and an increasing function of $a$ for $a>2\Omega_{R,0}/\Omega_{M,0}$. In addition, $P(a, t)$ is an increasing function of $a$ also for $a<0$.  
\end{lemma}

Manifestly, by setting $P(a,t)=0$ we recover Eq.~(\ref{eqRM}), a fact which allows us to find the desired solution $a (t)$.  Correspondingly, we are dealing with a third order algebraic equation. The fundamental theorem of algebra ensures that for each value of $t$ the equation $P(a,t)=0$ possesses three roots, which can be all real or one real and two complex conjugated since $P(a,t)$ has real coefficients. Owing to the physical meaning of $t$, we assume $t \geq 0$. From an algebraic point of view $P(a,t)=0$ possesses the three solutions for each value of $t$
\begin{eqnarray}
\label{a1}
a(t) = \frac{\Omega_{R,0}}{\Omega_{M,0}} \left[1-2 \, {\rm sin} \left(\frac{1}{3} \, {\rm arcsin} \, X(t) \right) \right]~,
\end{eqnarray}
\begin{eqnarray}
\label{a2}
a(t) = \frac{\Omega_{R,0}}{\Omega_{M,0}} \left[1 + 2 \, {\rm cos} \left(\frac{1}{3} \, {\rm arccos} \, X(t)  \right) \right]~,
\end{eqnarray}
\begin{eqnarray}
\label{a3}
a(t) = \frac{\Omega_{R,0}}{\Omega_{M,0}} \left\{1+2 \, {\rm cos} \left[\frac{1}{3} \, \bigl(2 \pi + 
{\rm arccos} \, X(t) \bigr) \right]\right\}~,
\end{eqnarray}
where we have set for notational simplicity 
\begin{equation}
\label{28122020a}
X(t) \equiv 1- 3 \frac{\Omega_{M,0}^2}{\Omega_{R,0}^{3/2}}H_0 t+ 
\frac{9}{8}\frac{\Omega_{M,0}^4}{\Omega_{R,0}^3} H_0^2 t^2~.
\end{equation}   

In order to understand which of them is physically meaningful, we perform an analytical study of the function $P(a,t)$. We stress that in order for ${\rm arcsin} \, X(t)$ and ${\rm arccos} \, X(t)$ to be real-valued we must have $- 1 \leq X(t) \leq 1$, but for $X(t) > 1$ ${\rm arcsin} \, X(t)$ is complex while ${\rm arccos} \, X(t)$ is imaginary. Explicitly 
\begin{equation}
\label{03012021a}
{\rm arcsin} \, X(t) = - i \, {\rm ln} \, \left(i X(t) + \sqrt{1 - X^2(t)} \right)~, 
\end{equation}
\begin{equation}
\label{03012021b}
{\rm arccos} \, X(t) = - i \, {\rm ln} \, \left(X(t) + \sqrt{X^2(t) - 1} \right)~. 
\end{equation}
 
We are now ready to accomplish our task, keeping in mind that $\Omega_{M,0}$, 
$\Omega_{R,0}$ and $H_0$ are real positive quantities.

We start by observing that from Eq.~(\ref{poly}) we have
\begin{equation}
\label{lim}
\lim_{a \to +\infty}P(a,t)=+\infty~,
\end{equation}
and in addition $P(0,t) < 0$ provided that $t$ satisfies the condition
\begin{equation}
\label{constTerm}
t > t_1
\end{equation}
with $t_1$ defined by
\begin{equation}
\label{t1def}
t_1 \equiv \frac{8\Omega_{R,0}^{3/2}}{3 \Omega_{M,0}^2 H_0}~.
\end{equation}
Assuming that Eq. (\ref{constTerm}) is met, owing to Eq.~(\ref{lim}) and $P(0,t) < 0$ we see that $P(a,t)$ obeys the Bolzano theorem, which entails that $P(a,t) = 0$ possesses {\it at least}  one root in the range $0 \leq a < \infty$. Moreover, by combining $P(0, t)<0$ with our Lemma it follows that $P(a, t) = 0$ possesses {\it only one} real root in the range $0 \leq a < \infty$. Further, Eq. (\ref{constTerm}) implies that $X(t) > 1$. Therefore, from Eq. (\ref{03012021a}) we find that ${\rm arcsin} \, X(t)$ is complex, and so its sine is complex either. So, Eq.~(\ref{a1}) is one of the two complex conjugated roots. Turning our attention to Eq. (\ref{a2}), from $X(t) > 1$ and Eq. (\ref{03012021b}) we get that ${\rm arccos} \, X(t)$ is imaginary, and so its cosine is real, since ${\rm cos} \, i \, x = {\rm cosh} \, x$ for any real number $x$. Thus, we see that Eq.~(\ref{a2}) is a real root. Finally, we address Eq. (\ref{a3}). By the same token, we see that the argument of the cosine is complex -- owing to the presence of $2 \pi$ -- so that this is the remaining complex conjugated root. In conclusion, for $t > t_1$ only Eq.~(\ref{a2}) is 
physically acceptable.

\

Let us next consider the opposite case, namely  
\begin{equation}
\label{constTerm1}
0 \leq t < t_1~.
\end{equation}
Accordingly, we now have $P(0,t)>0$, and since  
\begin{equation}
\label{lim1}
\lim_{a \to - \infty}P(a,t)= - \infty
\end{equation}
the Bolzano theorem ensures that $P(a,t)$ possesses at least one root in the range $- \infty < a \leq 0$. Which one? In order to settle this issue, it is of paramount importance to realize that condition~(\ref{constTerm1}) implies $-1 \leq X(t) \leq 1$. Whence $0 \leq {\rm arccos} \, X(t) \leq \pi$. Therefore, we have $2 \pi/3 \leq (2 \pi + {\rm arccos} \, X(t))/3 \leq \pi$, which means that $- \, 0.5 \leq {\rm cos} \, [(2 \pi + {\rm arccos} \, X(t))/3] \leq - \, 1$. 
Correspondingly, we find that it is Eq. (\ref{a3}) which has a real negative root: this is manifestly unphysical and will be discarded. What about the other two solutions? Because we presently have $P(0,t) >0$ -- recalling our Lemma -- we conclude that there exist two real and positive solutions {\it a priori} physically acceptable. These solutions are represented by Eqs.~(\ref{a1}) and~(\ref{a2}). And by taking the limit for $t \to 0$ of both of them, we discover that only Eq.~(\ref{a1}) has the correct behavior, namely $a \to 0$ for $t \to 0$. As a result, when either radiation or matter dominates, we discover that close enough to $t=0$ the solution of the Friedmann equation is given by Eq.~(\ref{a1}), while for $t$ satisfying condition~(\ref{constTerm}) the solution becomes Eq.~(\ref{a2}). When do these two solutions join? Because they must {\it soothly} join, there will be a time $t_*$ such that
\begin{equation}
\label{junct}
{\rm sin} \left(-\frac{1}{3} \, {\rm arcsin} \, X(t_*) \right) = {\rm cos} \left(\frac{1}{3} \, {\rm arccos} \, X(t_*) \right)
\end{equation}
where we have employed the fact that the sine is an odd function. By using the trigonometric  relation ${\rm cos} \, (\pi/2 - \alpha) = {\rm sin} \, \alpha$ we can equal the arguments of the cosines, and we are left with the condition $3\pi/2 + {\rm arcsin} \, (X(t_*) ) = {\rm arccos} \, (X(t_*))$, which can be simplified as ${\rm arcsin} \, (X(t_*)) = - \pi/2$ by using the trigonometric relation ${\rm arccos} \, y = \pi/2 - {\rm arcsin} \, y$. Thus, we obtain that condition~(\ref{junct}) is verified when $X(t_*)=-1$ and -- recalling Eq.~(\ref{28122020a}) --  we finally get
\begin{equation}
\label{tsol}
t_*= \frac{4 \Omega_{R,0}^{3/2}}{3 \Omega_{M,0}^2 H_0}~.
\end{equation}

\

What remains to be done is to consider the case $t=t_1$. Accordingly, it is easy to show that $P(a,t_1)=0$ possesses two coincident roots $a(t_1)=0$. They can be found by inserting Eq.~(\ref{t1def}) into Eq.~(\ref{28122020a}), thereby getting $X(t_1)=1$, which -- once further inserted into Eqs.~(\ref{a1}) and~(\ref{a3}) -- yields $a(t_1)=0$ in either case: these solutions are not physically acceptable because $a (t)$ must be a monotonically increasing function of $t$ starting from $t = 0$ and $t_1 \neq 0$. The case of Eq.~(\ref{a2}) is different, since it gives $a(t_1)=3 \Omega_{M,0}/\Omega_{R,0}$, which is the physically acceptable solution.

\

As a result, the final {\it unique physical} solution of Eq. (\ref{eqRM}) -- namely of Eq. (19) of the main text -- can be expressed as
\begin{equation}
\label{solRM}
a(t)=
\begin{cases}

\begin{aligned}[]
\frac{\Omega_{R,0}}{\Omega_{M,0}}\left\{1-2 \, {\rm sin}\left[\frac{1}{3} \, {\rm arcsin}\left( 1-3\frac{\Omega_{M,0}^2}{\Omega_{R,0}^{3/2}}H_0 t+\frac{9}{8}\frac{\Omega_{M,0}^4}{\Omega_{R,0}^3}H_0^2t^2 \right) \right]\right\}
\end{aligned}~, &   \,\, t \le t_*~,\\[8pt]

\\

\\

\begin{aligned}[]
 \frac{\Omega_{R,0}}{\Omega_{M,0}}\left\{1+2 \, {\rm cos}\left[\frac{1}{3} \, {\rm arccos}\left( 1-3\frac{\Omega_{M,0}^2}{\Omega_{R,0}^{3/2}}H_0 t+\frac{9}{8}\frac{\Omega_{M,0}^4}{\Omega_{R,0}^3}H_0^2 t^2 \right) \right]\right\}
\end{aligned}~, &   \,\, t_* < t \le t_s~,

\end{cases}
\end{equation}
where $t_s$ is defined in the main text. Thus, Eq.~(\ref{solRM}) represents the solution of the Friedmann equation when either radiation or matter is important.


\section{Alternative form of the Friedmann equations}

Here, we derive an alternative mathematical expression of Eq.~(\ref{a2}). By employing Eq.~(\ref{03012021b}), we find that for $t_1 < t \le t_s$, $a(t)$ can be rewritten as
\begin{eqnarray}
\label{sol1alt}
a(t) = \frac{\Omega_{R,0}}{\Omega_{M,0}}\left\{1+2 \, {\rm cosh} \left[\frac{1}{3} \, {\rm ln}\left(X(t) +  \sqrt{X^2(t)-1}\right)\right] \right\}~,
\end{eqnarray}
in order not to encounter the imaginary unit during the calculation. The quantity $X(t)$ is given by Eq.~(\ref{28122020a}). Note indeed that Eq.~(\ref{a2}) produces at the end real positive value for $a$ even if the imaginary part appears during the calculation, as 
previously discussed.

\newpage

\section{Enlarged Figures of the main text}

\begin{figure}[h]
\centering
\includegraphics[width=1.2\textwidth]{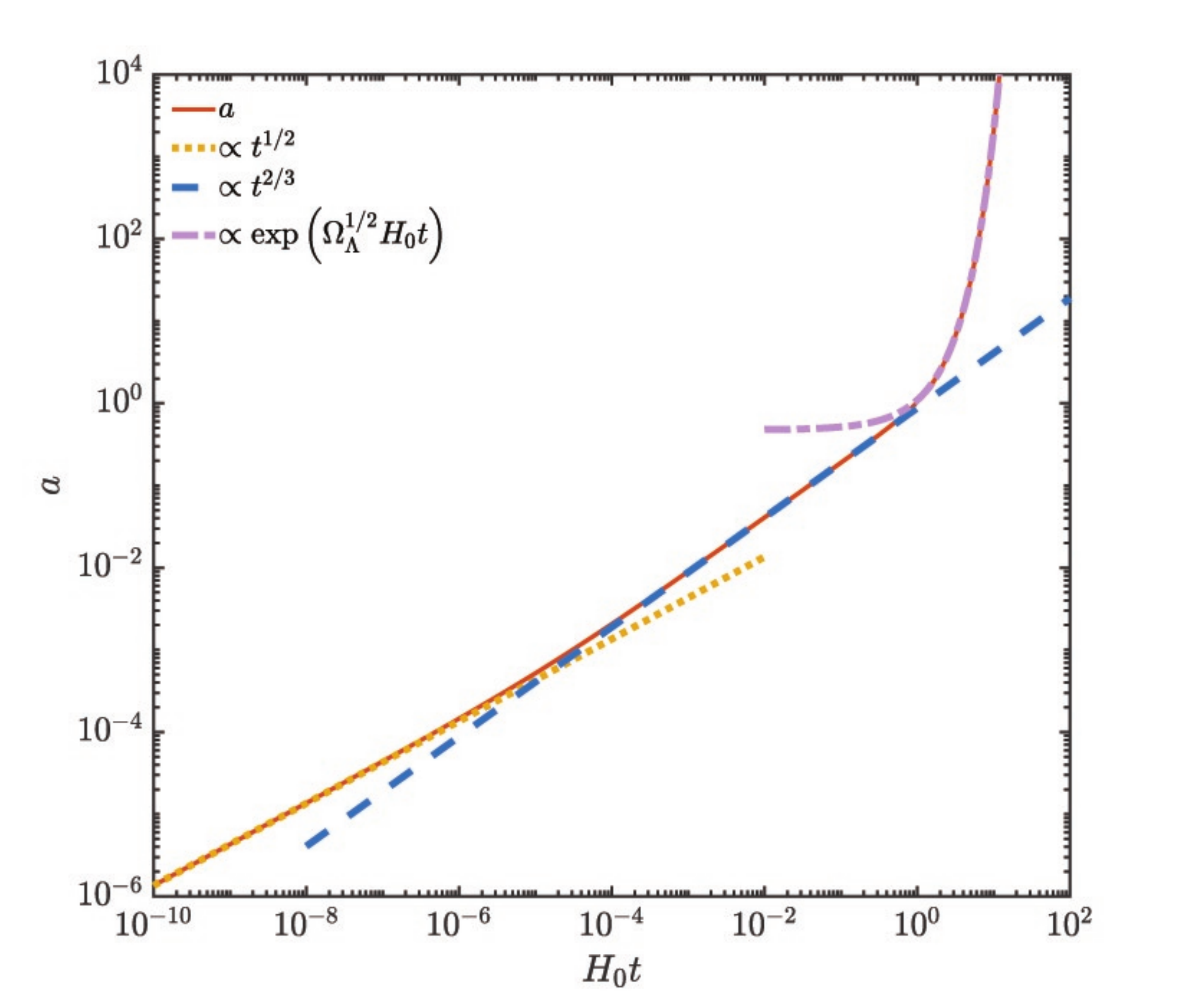}
\caption{\label{SF} Using our benchmark values, we plot the scale factor $a$ and the asymptotic behaviors versus $H_0t$.}
\end{figure}

\newpage

\begin{figure}[h]
\centering
\includegraphics[width=1.2\textwidth]{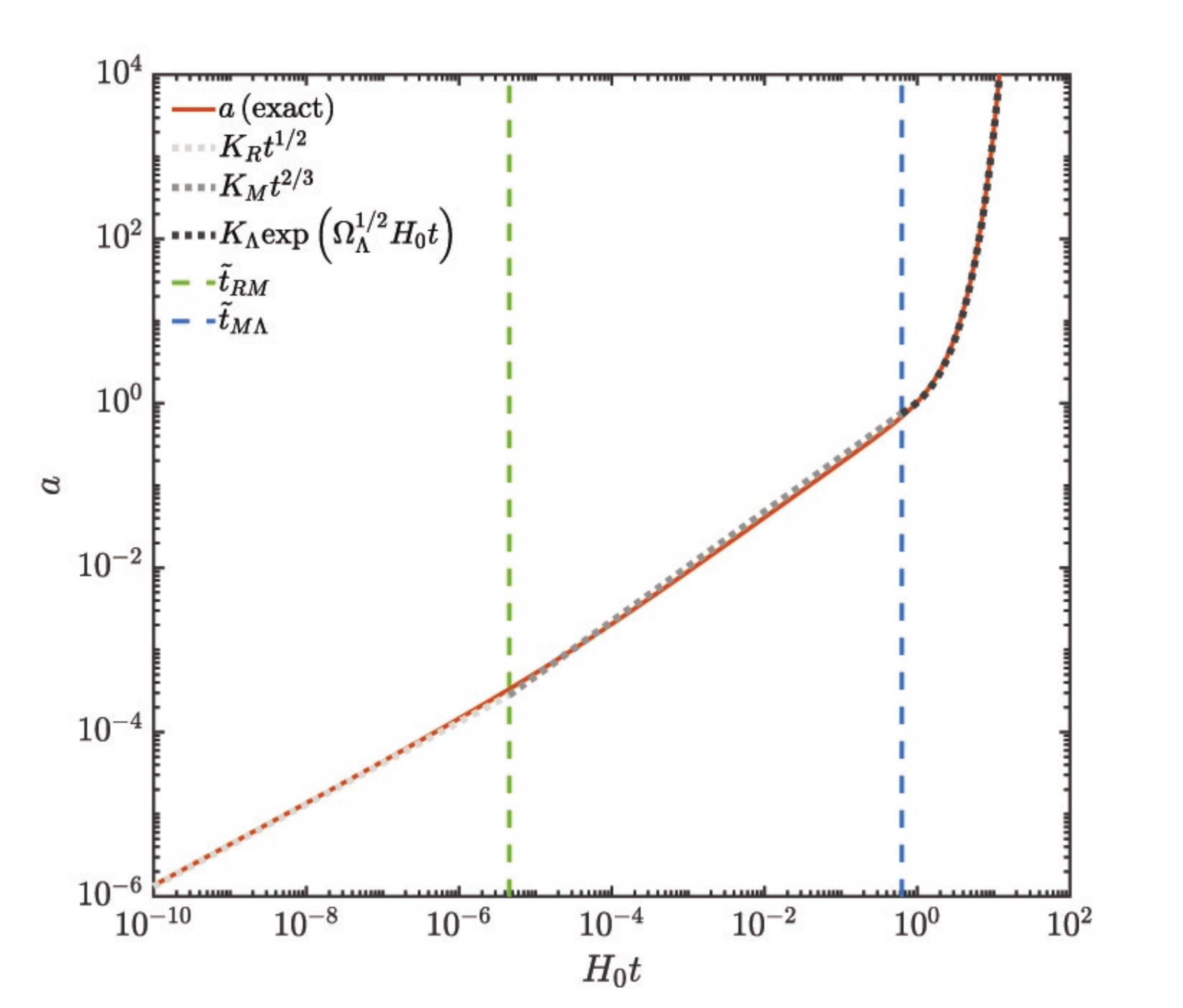}
\caption{\label{SFbroken} Exact scale factor $a$ and  piecewise function approximation $a_{\rm pw}$ as plotted versus $H_0t$.} 
\end{figure}

\end{widetext}


\begin{thebibliography}{20}

\bibitem{note1} As usual, we denote by $H_0$ the value of the Hubble constant, by 
$\Omega_{R,0}$, $\Omega_{M,0}$ and $\Omega_{\Lambda}$ the present values of the dimensionless cosmic density parameters pertaining to radiation, matter and cosmological constant, respectively. 

\bibitem{ryden} B. Ryden, {\it Introduction to Cosmology} (Cambridge University Press, Cambridge, 2017).

\bibitem{note2} We use $t_{R M}$ and $t_{M \Lambda}$ in connection with our exact solution. Below, in this Section we simply write $a (t)$ rather $a_{\rm pw} (t)$ for notational simplicity.

\bibitem{notaM} This can be seen multiplying Eq. (\ref{Friedmann}) by $a^2 (t)$, taking its time derivative and getting rid of $\ddot{a} (t)$ using Eq. (\ref{FriedmannII}).

\bibitem{note3} As benchmark values we take $\Omega_{R,0}= 8.47 \times 10^{-5}$, $\Omega_{M,0}=0.30$, $\Omega_{\Lambda}=0.70$, $H_0=70 \, \rm km \, s^{-1} \, Mpc^{-1}$, resulting in $t_0 = 13.47 \, {\rm Gyr}$. 

\end{thebibliography}
\end{document}